# Using Analogies to Learn Introductory Physics

Shih-Yin Lin and Chandralekha Singh

*Department of Physics and Astronomy, University of Pittsburgh, Pittsburgh, PA 15260, USA*

**Abstract.** Identifying the relevant physics principles is a central component of problem solving. A major goal of most introductory physics courses is to help students discern the deep similarities between problems based upon the physics principles so that they can transfer what they learned by solving one problem to solve another problem which involves the same principle. We conducted an investigation in which 251 calculus- and algebra-based introductory physics students were asked explicitly in the recitation quiz to learn from a solved problem and then solve another problem that has different surface features but the same underlying physics principles. We find that many students were able to discern the deep similarities between the problems. When the solved problem was provided, students were likely to invoke the correct principles; however, more scaffolding is needed to help students apply these principles correctly.

## INTRODUCTION

Learning physics is challenging. Physics is a subject which contains only a few fundamental principles that are condensed into compact mathematical forms. Learning physics requires unpacking these principles and understanding their applicability in a variety of contexts that share deep features [1-2]. A major goal of most calculus-based and algebra-based introductory physics courses is to help students learn to recognize the applicability of a physics principle in diverse situations and discern the deep similarities between the problems that share the same underlying physics principles but have different surface features.

One way to help students learn physics is via analogical reasoning [1-2]. Students can be explicitly taught to make an analogy between a solved problem and a new problem, even if the surface features of the problems are different. In doing so, students develop an important skill shared by experts: the ability to transfer from one context to another, based upon shared deep features.

Here, we examine introductory physics students' ability to perform analogical problem solving. Students were explicitly asked to point out the similarities between a solved problem and a quiz problem and then use the analogy to solve the quiz problem. In particular, students were asked in a recitation quiz to browse through and learn from a solved problem and then solve a quiz problem that has different surface features but the same underlying physics. Different types of scaffolding were provided in different intervention groups (recitation sections). The goal was to investigate what students are able to do with the analogy provided, and to understand if students could discern the similarities between the solved and the quiz problems, take advantage of them and transfer their learning to solve the quiz problem.

The solved problem provided was about a girl riding a roller coaster on a smooth track. The roller coaster car was initially at rest at a certain height. The problem asked for the apparent weight of the girl as the roller coaster car went over the top of a circular hump given the girl's weight, the radius of the circle, and the heights of different points. The quiz problem, on the other hand, was about a boy on a tire swing created with a rope tied to a tree. Students were asked to find the maximum tension in the rope during the ride given the boy's mass, the length of the rope, and the initial height assuming the boy starts from rest at the initial height. The same problems have also been used in other research [2]. Although these two problems look distinctly different in their surface features, both can be solved by first applying the principle of conservation of mechanical energy to find the speed at the point of interest, and secondly by applying Newton's $2^{nd}$ Law in the non-equilibrium with a centripetal acceleration involved. The roller coaster problem was provided to the students in the intervention groups with a detailed solution.

## METHODOLOGY

251 students from a calculus-based and an algebra-based introductory physics course participated in this study. Recitation classrooms for both courses were divided into 3 groups: comparison group, intervention 1, and intervention 2. Students in the comparison group received only the tire swing problem as a traditional quiz problem. They solved it on their own; no scaffolding was provided. The solved "roller coaster problem" was provided to students in the intervention groups 1 and 2. Our previous research indicates that simply giving students both the solved problem and quiz problem without additional scaffolding does not necessarily help students—many of them literally copy the equations from the solved problem to the quiz problem without contemplating their applicability. Different interventions in this study were therefore implemented with additional scaffolding with the expectation that the students will process the analogy more deeply.

**TABLE 1.** Summary of the rubric for the tire swing problem.

| Description | Correct answer | Common mistakes | Points taken off |
|---|---|---|---|
| Invoking and applying the principle of conservation of mechanical energy to find the speed (3 points) | $mg\Delta h = \frac{1}{2}mv^2$ | use kinematics equations to find v | 2 |
| | | wrong $\Delta h$ | 1 |
| Identifying the centripetal acceleration and use Newton's 2$^{nd}$ Law to find the tension (7 points) | $a = a_c = \frac{v^2}{r}$ $T - mg = ma_c = m\frac{v^2}{r}$ $\Rightarrow T = mg + m\frac{v^2}{r}$ | $a = 0, \quad T = mg$ | 5 |
| | | $a \neq 0$, but wrong formula for $a$ | 1~3 |
| | | $T = m\frac{v^2}{r}$ | 3 |
| | | $T = mg - m\frac{v^2}{r}$ | 2 |

In a recitation quiz, students in the intervention group 1 were first asked to browse over and learn from the solution to the roller coaster problem. They were explicitly told that after 10 minutes, they had to turn in the solution, and then they would be given two quiz problems to solve: one of them would be exactly the same as the solved problem they just browsed over (the roller coaster problem) and the other problem would be similar (the tire swing problem). By reproducing the roller coaster problem, students had the opportunity to think through the principles involved in more depth. We hypothesized that by asking students to solve the roller coaster problem after learning from and then returning its solution that was provided, they might pay more attention not only to the principles invoked, but also why and how each principle is applied while reading the solution. This additional attention may be advantageous for transfer.

Intervention 2 was designed based on our hypothesis that if the students struggled with the quiz problem first and were then provided the help of a solved problem (which involves the same physics principles), they would benefit more from it than if they were simply given the solved problem as an analogy to solve the quiz problem. We believed that having tried to solve the quiz problem, students would be more aware of their own difficulties and be more directed when learning from the solved problem. The struggling experience could give them a clearer picture about what the task is, and they could be more focused in terms of what to look for in the solved example. Intervention 2 was devised to fit this framework. Students in this group were first asked to solve the tire swing problem on their own. After 10 minutes, they turned in their first solution, and they were then given the roller coaster problem along with its solution. With the solved roller coaster problem in their possession, they were then given an opportunity to solve the tire swing problem a second time.

Students' performance on the quiz was later graded using a rubric. Two researchers discussed the rubric and scored independently a sample of 20 students. An inter-rater reliability of at least 80% was achieved. The rubric has a full score of 10 points, with 3 points devoted to the conservation of mechanical energy part, and 7 points for identifying the centripetal acceleration and all relevant forces, and applying Newton's 2$^{nd}$ law to obtain the final answer. A summary of the rubric highlights for the tire swing problem is shown in Table 1. The rubric for the roller coaster problem is similar. Based on the rubric, an analysis of students' performance in each group was performed and compared. To examine the effects of interventions on students with different expertise and to evaluate if the interventions were more successful in helping students at a particular level of expertise, students in each course were further classified as top, middle, bottom, and "none" by listing them in order based on their scores on the final exam and then splitting them into thirds (students in the category "none" didn't take the final exam). We investigated how top, middle and bottom students within the same intervention group performed with the same scaffolding provided; the effects of the interventions were also compared not only for the whole class taken together but also for each subgroup of students with a particular level of expertise.

# RESULTS

**TABLE 2.** Students' average scores out of 10 on the tire swing problem in the calculus-based course. The number of students in each case is shown in parentheses.

|  | Comparison | Intervention 1 | Intervention 2 |
|---|---|---|---|
| Top | 8.6 (14) | 9.3 (15) | 9.2 (13) |
| Middle | 7.6 (10) | 8.8 (10) | 9.4 (12) |
| Bottom | 4.2 (14) | 4.6 (11) | 8.7 ( 9 ) |
| None |  |  | 7.5 ( 4 ) |
| All | 6.7 (38) | 7.8 (36) | 9.0 (38) |

**TABLE 3.** Students' average scores out of 10 on the tire swing problem in the algebra-based course. The number of students in each case is shown in parentheses.

|  | Comparison | Intervention 1 | Intervention 2 |
|---|---|---|---|
| Top | 6.0 (19) | 8.0 (10) | 6.8 (12) |
| Middle | 2.7 (15) | 7.3 (20) | 6.7 (10) |
| Bottom | 2.0 (20) | 6.6 (16) | 4.8 (11) |
| None | 1.0 ( 2 ) | 6.0 ( 2 ) | 7.5 ( 2 ) |
| All | 3.5 (56) | 7.2 (48) | 6.2 (35) |

Tables 2 and 3 show the average scores on the tire swing problem for each group of students in the calculus-based and algebra-based courses, respectively. For intervention 2, in which the students had the opportunity to solve the quiz problem twice, the later scores when the students re-solved the problem with scaffolding are presented in Table 2 and 3. Intervention 2 students in both courses and intervention 1 students in the algebra-based course on average scored significantly higher (with $p$ values less than 0.05) than those in the comparison group, indicating that these students, to a moderate extent, could reason about the similarities between the two problems and take advantage of the solved problem provided to solve the quiz problem. Intervention 1 students in the calculus-based course, on the other hand, didn't show a significant improvement on average as compared to the comparison group. A possible explanation is that the bottom students in this group failed to process through the solutions deeply and they didn't perform well on the quiz problem as compared to other students who received the same intervention. We will discuss this in more detail in later paragraphs. Table 2 and 3 also show that the calculus-based students on average performed better than the algebra-based students in all three groups, regardless of the scaffoldings they received. The difference between the score of the calculus- and algebra-based students decreaseed after the interventions were implemented.

Students' common difficulties on the tire swing problem when no scaffolding was provided included: missing the gravitational force term, failing to recognize the fact that there was a centripetal acceleration and treating the problem as an equilibrium situation with only the gravitational force acting on the rope, realizing the presence of the acceleration but not knowing how to find it, and a difficulty in finding correctly the speed at the point of interest. Some students improperly used the kinematics equations instead of the principle of conservation of mechanical energy to find the speed; others erroneously associated the centripetal acceleration with the gravitational acceleration and used the equation: $v^2/r=g$ to solve for $v$ (here $v$ stands for the speed, $r$ is the radius of the circle, and $g$ is the gravitational acceleration). These mistakes were reduced when the scaffolding with the solved roller coaster problem was provided. With the scaffolding, more students were able to identify the existence of both the gravitational force and the centripetal acceleration, and most students could apply the principle of conservation of mechanical energy to find the speed correctly.

**TABLE 4.** Average scores out of 10 on the tire swing problem before and after the scaffolding for both the algebra-based and calculus-based students in intervention 2.

|  | Calculus-based | | | Algebra-based | | |
|---|---|---|---|---|---|---|
|  | before | after | gain (%) | before | After | Gain (%) |
| Top | 6.2 | 9.2 | 80 | 2.8 | 6.8 | 55 |
| Middle | 4.9 | 9.4 | 89 | 0.9 | 6.7 | 64 |
| Bottom | 4.7 | 8.7 | 75 | 1.7 | 4.8 | 37 |
| None | 2.0 | 7.5 | 69 | 5.5 | 7.5 | 44 |
| All | 5.0 | 9.0 | 79 | 2.1 | 6.2 | 52 |

Table 4 presents the consecutive scores of students in intervention 2 on the tire swing problem before and after browsing over the solved problem. Normalized gains [3] of 79% and 52% on average were found for the calculus- and algebra-based students, respectively. It appears that the intervention worked very well for the calculus-based students; even the bottom students achieved an average score of 8.7 out of 10. Comparing students' work before and after the scaffolding shows that the significant improvement is due to the fact that most calculus-based students were able to correctly invoke the necessary knowledge which they lacked initially and corrected at least part of their own mistakes after browsing over the solution.

Table 5 presents students' performance on the roller coaster problem in intervention 1 after they browsed over and then returned the solution to the same problem. The performance on their quiz problem is also listed for comparison. Table 5 shows that many students in both the calculus- and algebra-based courses were capable of reproducing the solved problem to a considerable degree immediately after browsing over (and returning) the solution. The average scores of the "bottom" students on the solved problem reproduced were 8.2 (calculus) and 8.5 (algebra); the scores of the middle and top students were even higher. These scores, however, could be superficial with regard to the students' ability to transfer the learning to a new problem. Indeed, the score on the quiz problem dropped to 4.6 for the "bottom" calculus-based students. An average drop of 1.7 points was also found for the algebra-based students. In general, the top and middle students in the calculus-based course were better at transferring their learning from the solved problem to the quiz problem although the algebra-based students had higher scores when reproducing the solved problem. We believe that students' ability to learn and transfer what they learned from one problem to another is dependent on their existing knowledge structure. The intervention may be successful only if the scaffolding is commensurate with students' current knowledge [4] and is not beyond the students' zone of proximal development. Our previous research indicates that the roller coaster problem and the tire swing problem are quite difficult for the calculus-based students, and the algebra-based students have considerably more difficulty. Even though the students in the algebra-based courses are in general strongly motivated to learn because they are typically interested in careers in health professions, their initial knowledge could have a strong impact on how much they could gain from the scaffolding provided. On the other hand, the "bottom" students in the calculus-based group might not have as strong a motivation to perform well as the algebra-based students, and that could be a possible reason for why their score was on average lower than other subgroups which received the same scaffolding (see Table 5).

**TABLE 5.** Average scores out of 10 on the roller coaster problem (solved problem) and the tire swing problem (quiz problem) for intervention 1 in algebra- and calculus-based courses.

|  | Solved Problem | | Quiz Problem | |
| --- | --- | --- | --- | --- |
|  | Calculus | Algebra | Calculus | Algebra |
| Top | 9.0 | 9.6 | 9.3 | 8.0 |
| Middle | 8.7 | 9.0 | 8.8 | 7.3 |
| Bottom | 8.2 | 8.5 | 4.6 | 6.6 |
| None |  | 8.0 |  | 6.0 |
| All | 8.7 | 8.9 | 7.8 | 7.2 |

Although students were in general able to take advantage of the roller coaster problem and identify the relevant principles involved in the tire swing problem, some students failed to apply the principles correctly, e.g., they failed to note how the new situation required changes in the details of applying the principles. For example, for the roller coaster problem, the cart was going over a hump at the point of interest, and the centripetal acceleration was pointing down. On the contrary, the maximum tension in the tire swing problem occured at the bottom of the circle, with the centripetal acceleration pointing up. Some students failed to differentiate between these two cases; they used the same equation that was used for the normal force in the solved problem to solve for the maximum tension in the tire swing problem and arrived at an expression that has an incorrect sign for the centripetal acceleration: $T=mg-mv^2/r$. This answer for the tension force was rarely observed if students were not provided any scaffolding. This mistake was more common in the algebra-based course than in the calculus-based course. Examination of students' work indicates that many students didn't draw a free body diagram when solving the problem. It is possible that these kinds of mistakes could be reduced if, in addition to the current intervention, students are explicitly asked to draw a free body diagram before solving the problem, and a comparison between the free body diagrams for the tire swing problem and the roller coaster problem is explicitly enforced.

## DISCUSSION

Many students in both the algebra-based and calculus-based courses were able to discern the similarities between the solved problem and the quiz problem and were able to immediately transfer what they learned from the solved problem provided to solve the quiz problem. Students who received the solved problem typically did well in invoking the correct principles when solving the quiz problem; however, the application of the principles was often challenging for them. More scaffolding may help students discern the similarities and differences between the solved and quiz problems in order to apply the principles correctly. It will also be interesting to see if students are able to invoke and apply the principles they learn in this activity when later in an exam they are asked to solve a problem involving the same principles. In all, we believe these kinds of activities are beneficial in helping students understand the applicability of physics principles in diverse situations and discern the coherence of the knowledge structure in physics. The greatest benefit may be achieved if similar activities are sustained throughout the course and the coherence of physics and the importance of looking at the deep features of the problems is consistently emphasized, demonstrated and rewarded by the instructors.